# A solution for actors' viewpoints representation with collaborative product development

Hichem M. Geryville [1], Abdelaziz Bouras [1], Yacine Ouzrout [1], Nikolaos S. Sapidis [2]

**(1) :** Prisma Laboratory – Lyon 2 team
IUT Lumière Lyon 2
160 Boulevard de l'université 69676
Bron cedex, France
*+33(0)4.78.77.24.36/+33(0)4.78.00.63.28*
*E-mail* : {hichem.geryville, abdelaziz.bouras, yacine.ouzrout}@univ-lyon2.fr

**(2) :** Department of Product and Systems
Design Engineering
University of the Aegean
Ermoupolis – Syros GR-84100, Greece
*+30.228.10.97.110/+30.228.10.97.009*
*E-mail* : sapidis@syros.aegean.gr

**Abstract:** As product complexity and marketing competition increase, a collaborative product development is necessary for companies which develop high quality products in short lead-times. To support product actors from different fields, disciplines, and locations, wishing to exchange and share information, the representation of the actors' viewpoints is the underlying requirement of the collaborative product development. The actors' viewpoints approach was designed to provide an organisational framework following the actors' perspectives in the collaboration, and their relationships, could be explicitly gathered and formatted. The approach acknowledges the inevitability of multiple integration of product information as different views, promotes gathering of actors' interests, and encourages retrieved adequate information while providing support for integration through PLM and/or SCM collaboration. In this paper, a solution for neutral viewpoints representation is proposed. The product, process, and organisation information models are seriatim discussed. A series of issues referring to the viewpoints representation are discussed in detail. Based on XML standard, taking cyclone vessel as an example, an application case of part of product information modelling is stated.

**Key words**: Viewpoints, multidisciplinary collaboration, exchange information, product and process information, product lifecycle management.

## 1- Introduction

The development of a successful product is achieved through the cooperation of various actors' teams and utilization of an amount of different resources existing in various disciplines, organisation and locations. As the complexity of products and market competition increase, a collaborative product development framework is necessary for product engineers developing high-quality products in short lead-times [BW1, VT1]. In this logic, various actors coming from different fields, disciplines, and locations are required to collaboratively develop a product efficiently. To support this collaborative work process, it is necessary to exchange and share information among product developers. The exchanged and shared information must be understood by various actors without any ambiguity even if they come from different industries or disciplines. Hence, common, neutral viewpoint/knowledge representation is required to help actors to capture/retrieve the adequate exchanged/shared information following their perspectives/objectives on the collaboration, because each actor depends on his own experiences/knowledge and interpret differently the same information on the product. To apply it, a universal information models are necessary to assure all actors within the whole collaborative product development process can clearly understand it.

In multidisciplinary collaboration, the actors express their interests (activity's focus) using a variety of conditions and/or representations, and follow different processes to deploy and extract those representations. Such multi-perspective product development embodies a number of characteristics that continue to be positioned in the core of organizing the product development activity. These characteristics include the need for gathering of actors' interests during product lifecycle phases following their points of view. The multiple comprehension of the information of processes and products, and the need to reason analytically over multiple views, permit to understand the properties and consequences of the multiple-viewpoint definitions of the actors. In other words, the viewpoints approach must define the specific information needed by each actor following his interests on the product/process.

To assure that the collaborative framework can be utilized by all actors, it must be based in two important layers: i) neutral base-level information that is independent of any application from respective collaborators, and ii) representation of the actors' viewpoints which represent their knowledge and can help them to retrieve the adequate information following their personal objectives in the collaboration.

In this paper, we focus on proposing the multiple viewpoints approach to organise and manage the exchange and sharing of product/process information between actors. The paper reviews two main viewpoint definitions (viewpoints on systems' specification, product design views). The paper presents the product-process-collaboration-organisation model as base-level modelling. The paper describes the use of explicit relationships between viewpoints to manage information extraction on the product development. An example has been given to illustrate the proposed approach.

## 2- Literature review

In the literature, a lot of studies focusing on knowledge representation and product modelling during collaborative development process have been investigated [K1, ST1, ZL1]. However, most of them emphatically address only a part of the entire issue of product modelling. And, up until now, few papers have synthetically investigated product information modelling based on the users' points of view on a large-system, from an application system in which the product information models are used to description of product information models.

Based-on existing literature, we can state that is difficult to find a general definition of "viewpoints" in the collaborative product development area; however, there are a lot of definitions related to the Requirements Engineering (RE) and Concurrent Engineering (CE).

One of the known viewpoints framework is the Reference Model Open Distributed Processing (RM-ODP) [I2]. The RM-ODP defines a holistic framework for the specification of all distributed systems. In order to deal with all aspects and complexities of such systems, the reference model defines five different abstractions – referred to as viewpoints – from which distributed systems may be modelled: *information, organizational, computational, engineering, and technology*. These viewpoints are sufficiently independent to simplify reasoning about the complete specification of the system; they are also generic and complementary. They enable different abstraction viewpoints, allowing actors to observe a system from different suitable perspectives [I3]. One of the main benefits of RM-ODP is that, as opposed to other approaches – as IEEE standard 1471 [I1], the "4+1" view model [K2], or the Zachman's framework [Z1] – it provides precise definitions of a system of interrelated concepts rather than some, often imprecise, description of isolated ones.

Among ODP viewpoints, we are here interested in the information viewpoint which is related to the information modelling. An information specification defines the semantics of information and information processing in an ODP system, without having to worry about other system considerations, such as particular details of its implementation.

Ribière [R1] considers "Viewpoint" as a polysemous word, i.e. its definition depends on the context of use. She defines a viewpoint as "*a perspective of interest from which an expert examines the knowledge base*". It is a general definition that can take several interpretations in different domains of application. She proposes an extension of the conceptual graph formalism to integrate viewpoints in the support and in the building of conceptual graphs. Where the viewpoints allow her to define the context of use and the origin actor of concept types introduced in a graph. The aim of her proposal is to define viewpoints to help knowledge representation with conceptual graphs for multi-expert knowledge acquisition and also to have an accessible and evolutive knowledge base of conceptual graphs through viewpoints.

Other viewpoints definitions have been given for Garlan [G1] *a viewpoint can be defined as a simplifying abstraction of a complex structure… suppressing information not relevant to the current focus;* and for Easterbrook [E1] *a viewpoint represents the context in which a role is performed.*

From the product views side, Hoffman [HJ1] proposes a mechanism for maintaining consistent product views in a distributed product information database. In his work, a single repository called a "master model" in which all-relevant product data resides was proposed for the integration of different product information domains, while the other views of the product must be updated to maintain consistency. Thus, Bronsvoort [BN1] proposed a multiple viewpoints feature modelling approach to allow a designer to focus on the information that is relevant for a particular product development phase. His system supports conceptual and assembly design, part detail design and part manufacturing planning by providing their own view on a product for each of these applications.

The scope of the presented works mainly deals with the integration of a domain point of view (e.g. design) for all actors, and the generation of the product views according to the actor's objective. However their limitation is that none of them integrate the interests of the actors within different product lifecycle stages, such as the extraction/retrieve of adequate information. In multidisciplinary collaboration, the actors need to integrate their viewpoints on the product along its lifecycle stages, where they also need to retrieve important information within different stages according to their interests.

Among the previous definitions, we think that Ribière [R1] and Easterbrook [E1] are nearing adapted to the product lifecycle collaboration area. However, we need to improve viewpoint solution by adopting the adaptation of actors' knowledge/information within product lifecycle and supply-chain collaboration.

Based on previous studies on issues of collaborative product development process [GB1, GO1], this paper attempts to contribute to defining a solution for actors' viewpoints representation, where we propose an approach that makes connections with product, process, and organization information for a complete interaction between multidisciplinary actors. To give an appropriate definition of multiple viewpoints actors, we situate the actors under different views (product, process and collaborative supply chain organization). In the next section, we define our notion of viewpoints and its interactions with the product/process information within a collaborative supply chain organization.

## 3- Proposed approach

In multidisciplinary collaboration, the framework must be characterized with following features:

- a base-level information model should contain enough information for various needs of the product

collaborators, i.e., the product information model should be based on the whole lifecycle of the product to assure it contains complete information meeting the various demands of all actors in all product development stages.

- actors' viewpoint should represent the actors' knowledge and interests on the product collaboration which permit to help them to seek, extract, exchange, and share the adequate information. In a word, a whole-life-cycle product information model with neutral viewpoint representation is necessary to support collaborative product development.

### 3.1 – The information models

In a previous study [GB1, GO1], we defined an information model called PPCO (Product–Process–Collaboration–Organization information model) based, especially, on the collaborative and project options, this study is inspired 50% from [GR1] and [NR1] where they propose the Product–Process–Organization model (PPO). The PPCO provides a base-level information model that is open, extensible, independent from any product development process, aiming at capturing the engineering and business context commonly shared in product development (Fig. 1.).

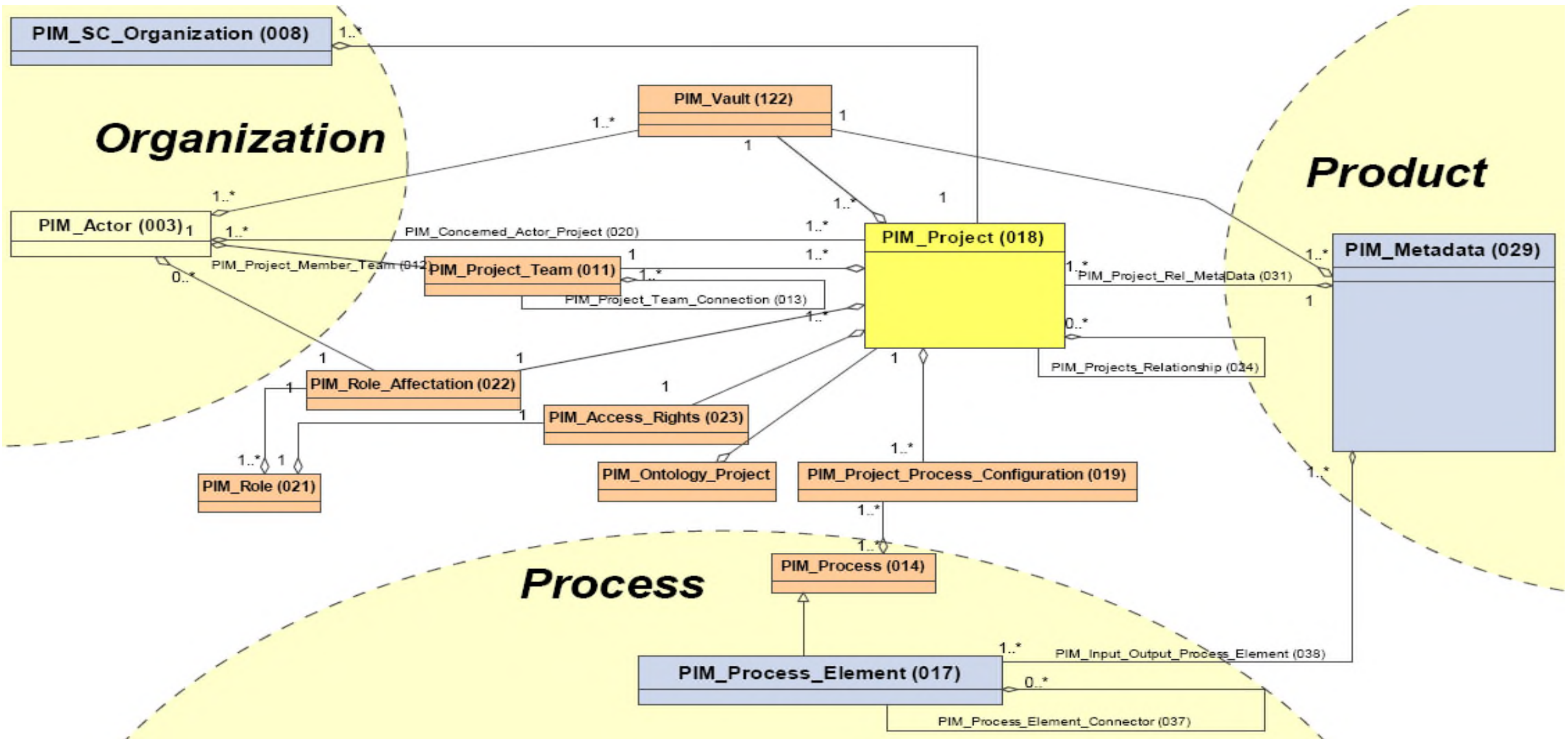

**Figure 1: The PPCO Model.**

The PPCO model is based on the four elements stated previously: product, process, collaboration and organization.

- **Product:** the architecture of the product is defined not only by the decomposition of the final product into components, functions, behaviours, etc, but also by the interactions between all these components. The interactions may include well-specified interfaces and undesired or incidental interactions (Fig. 2).
- **Organization:** the organization structure determines who works with whom and who reports to whom. However, in supply chains organization we are particularly interested in the study of the communication patterns between the actors conducting the technical development work, and the classification of the actor's competences following the integrated collaboration's domains (Fig. 3).

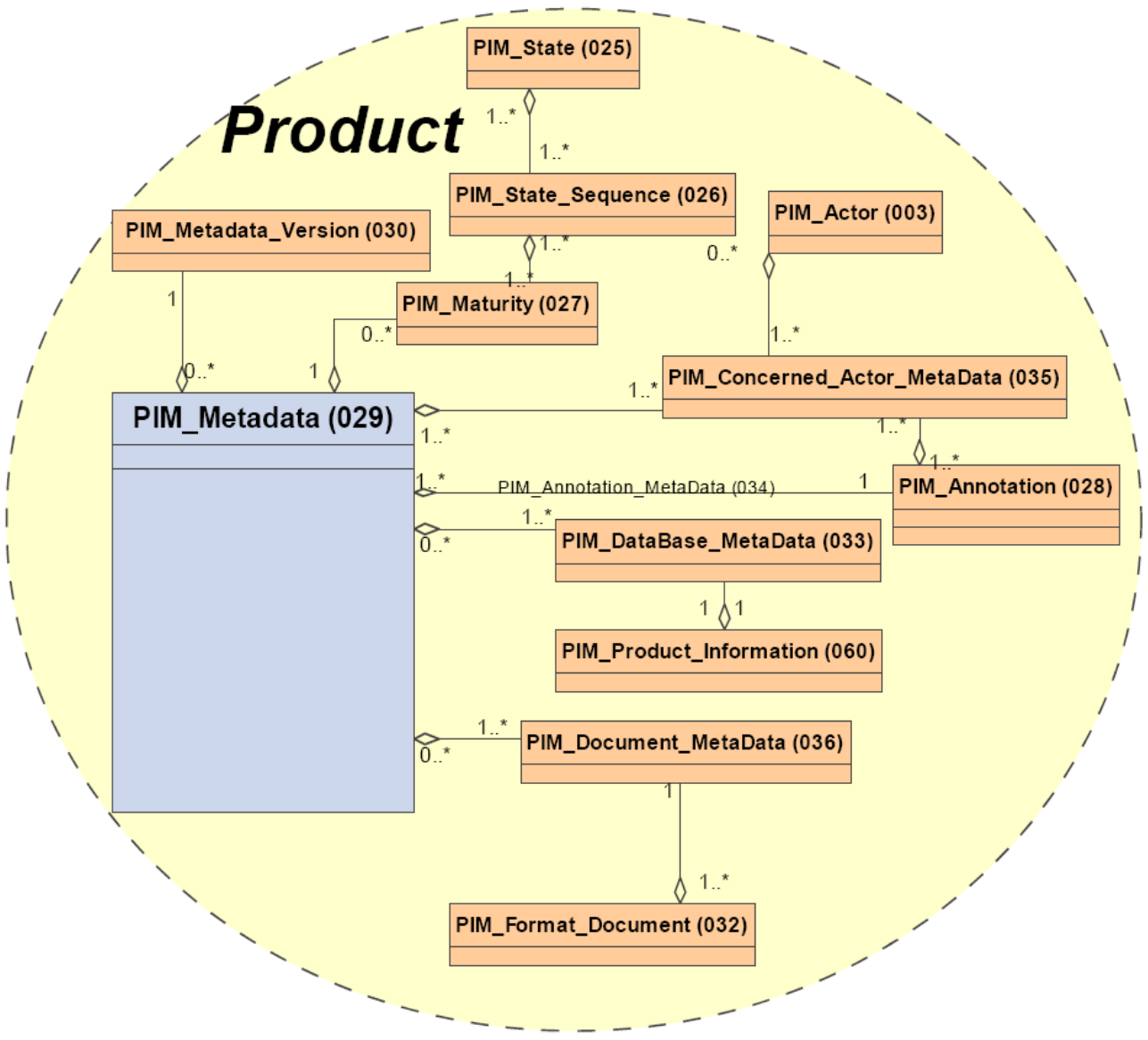

**Figure 2: The Product Model.**

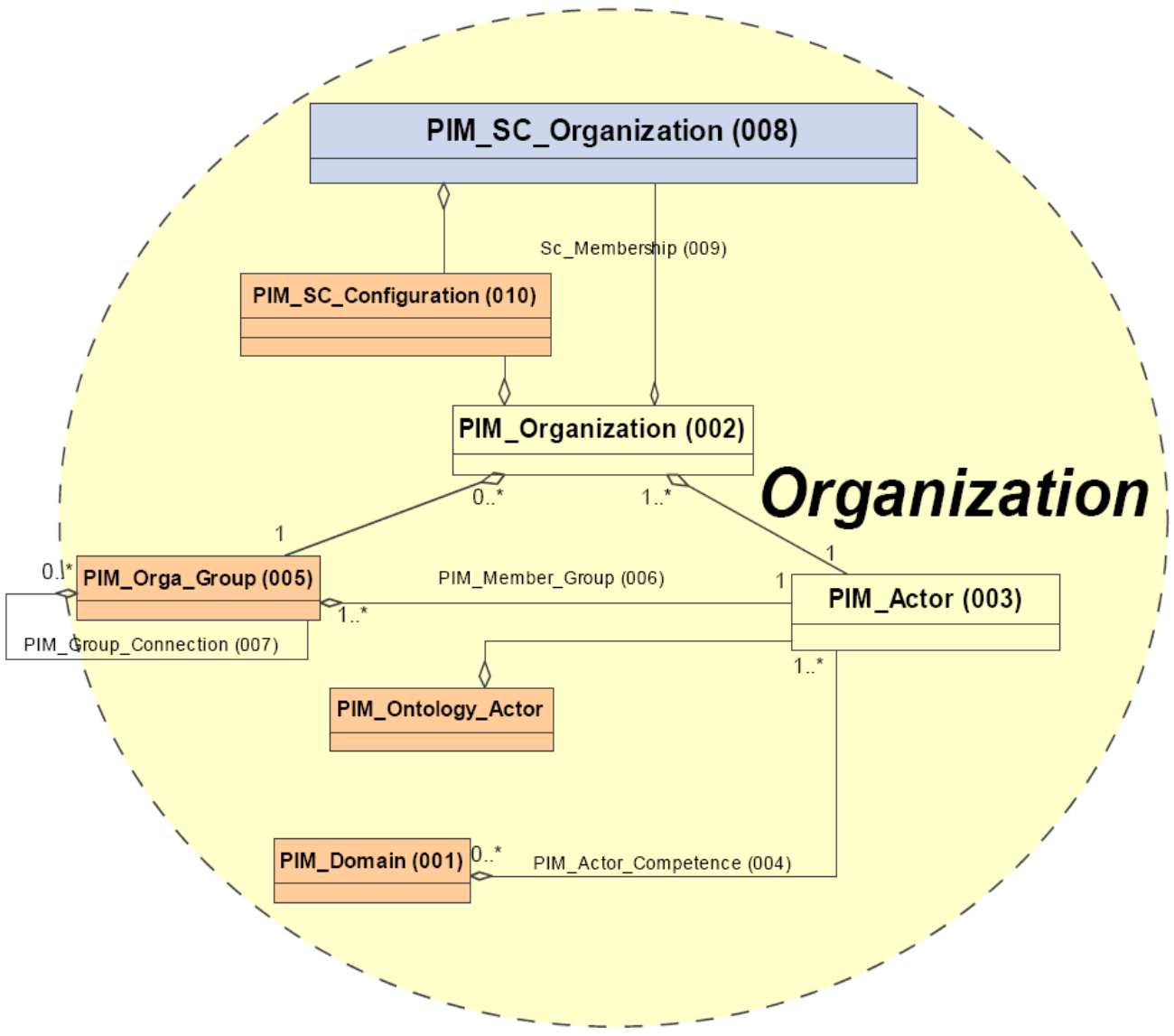

**Figure 3: The Organization Model.**

- **<u>Process:</u>** the product development process is generally a complex procedure involving information exchange across the many activities/tasks in order to execute the collaborative work. Various network-based methods have been used to map and study development processes (Fig. 4).

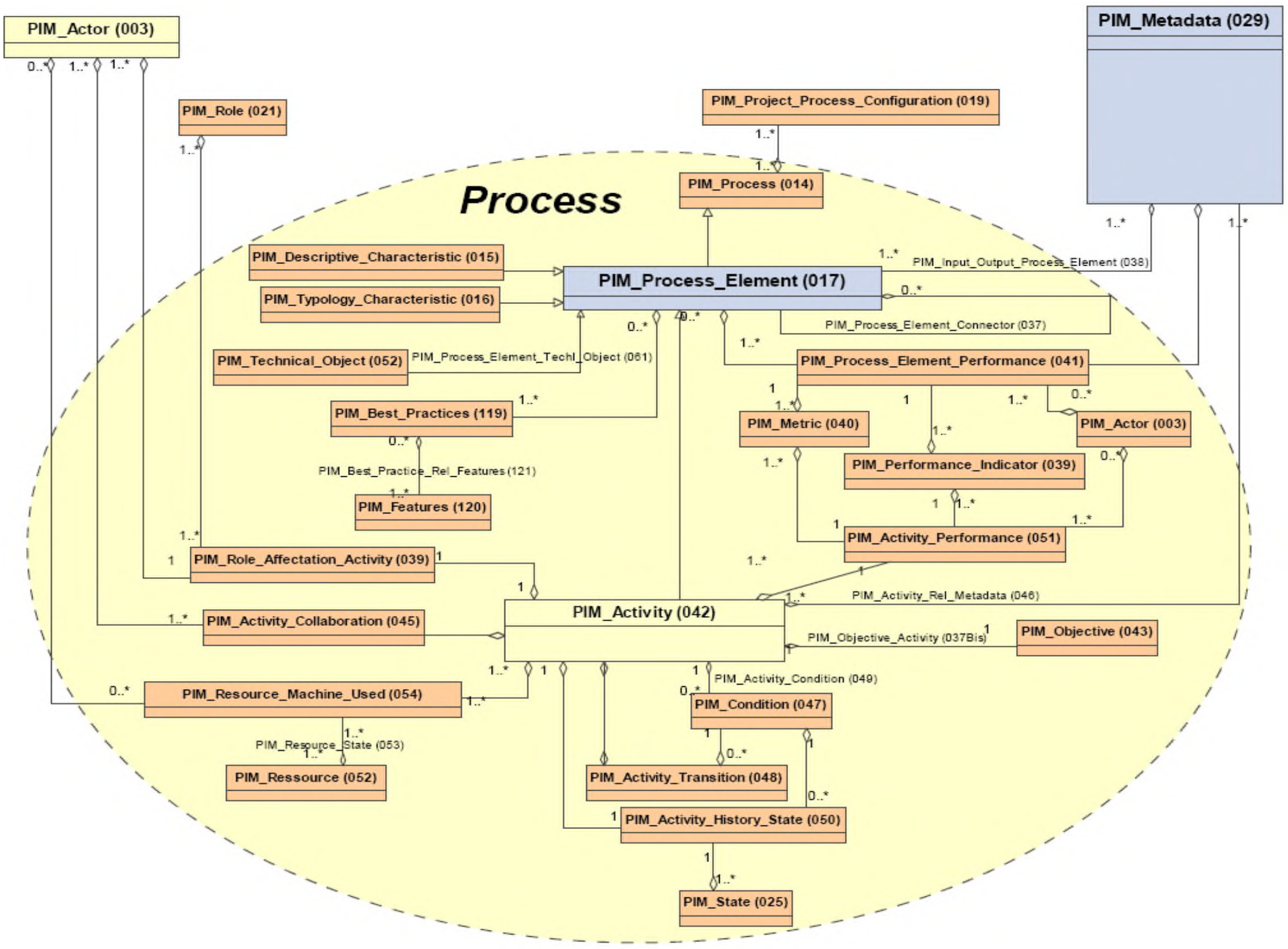

**Figure 4: The Process Model.**

### 3.2 – The multiple viewpoints approach

Our notion of a viewpoint must take into account that the viewpoint is an object encapsulating cross-cutting and partial knowledge about activity, process, and domain of discourse, from the perspective of a particular actor, or collaboration-team, in the product development process.

Fundamentally, viewpoints organize the knowledge of product/software development based on gathering of interests. A viewpoint, expresses the focuses of a particular actor, such as a developer or a representative of an area of interest captured by that viewpoint. Thus, a viewpoint may represent an area of interest within a project, a product, or a process, or may simply present a particular perspective expressed in a particular notation.

The degree of importance and reuse of information changes from one actor to another according to his objectives and activities related to the product. The use of this information depends on different actors' viewpoints, competencies, skills, responsibilities and interests on product information and product lifecycle phases. Several actors use the product information differently according to the specific requirements of their discipline.

#### 3.1.1 –Approach description

In a multidisciplinary collaboration (MC) context, the human dimension is important and corresponds to the different actors in a product lifecycle phases, the related knowledge dimension corresponds to the experience, competence and situation of the actors. Most of the authors agree that a viewpoint is strongly influenced by the domain area of the actor. Factors such as the field of expertise and specific technical interest play a role in this representation. Several actors see the product differently according to the constraints specific to their discipline.

Based on both definition of Garlan [G1] and Easterbrook [E1], we a define viewpoint as *a subset of information concerning the description of a product respecting the actor's focuses.* This definition is characterized by a context, which allows the restitution of the information that the actor want to use/retrieve, and a degree of importance/reuse he wants to give to this viewpoint.

In our case, the viewpoint permits:

- Simple seek of information of product/process within a supply chain context,
- Visualization of pieces of information according to a given process/activity,
- Comparison of information between viewpoints.

So, in a MC context we notice that each actor has one or more viewpoints following his activities in different phases of the product lifecycle or supply chain process. A viewpoint is characterised by four objects (or concept types): *i)* the **Actor** concept is described by the actor's information, and his situation, which defines his role through the collaboration, and his competence level on specific domain/discipline; *ii)* the **Viewpoint_Domain** concept is defined by the current actor activity in the product development process which is dependent on the actor's situation; *iii)* the **Viewpoint_Relationship** concept indicates the different interactions between actor's viewpoints, where the viewpoint information will be compared and gathered in one global information-set, and will be used for filtering actions; *iv)* the **Viewpoint_Objective** concept is described by a focus which defines the actor's objective according to it's activity objective and the viewpoint domain of the objective's activity. Figure 5 shows the different relationships between the four concepts of viewpoint.

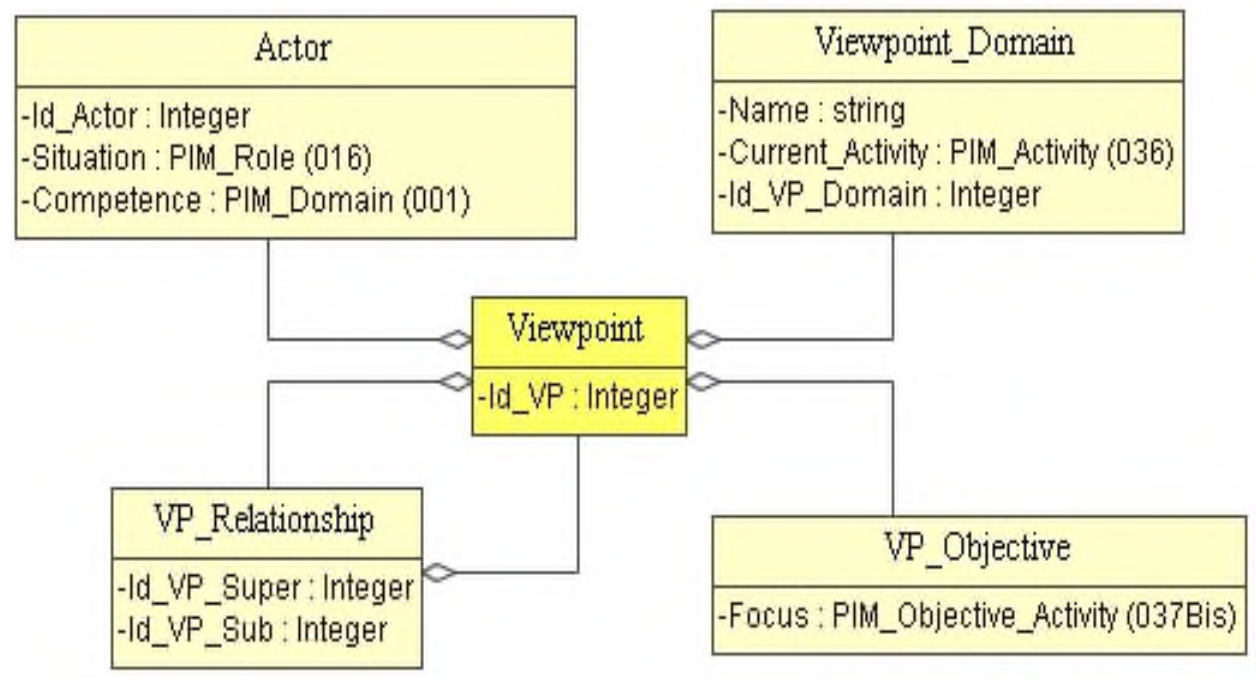

**Figure 5: Viewpoint definition.**

### 3.1.2 –Algorithm for the criteria filtering

In the table 1, we present one of the main information filtering algorithms. The filtering action consists in retrieving the appropriate information for a particular actor, according to his "viewpoint" on the product. Also it permits to combine all viewpoints and adapt the found information. However, the actors can carry out modifications, according to their rights, on the recorded information. For each modification, the actor submits his update which will be temporarily saved in the database of the PPCO model, and annotated to all actors concerned by this change. This update will take effect after approval of these concerned actors.

```
Program filtering_info_artifact (In: Artifact#, Actor#;
                                 Out: List_connexion_batch_info)
Begin
  // 1st step: Restitution of all actor's viewpoints
  Restitution_list_viewpoint(In: Actor#; Out: List_vp_actor);
  // 2nd step: Filtering of viewpoint on current artifact
  Filtering_list_vp_artifact(In: List_vp_actor, Artifact#; Out: List_vp_actor_prod);
  // 3rd step: Viewpoints classification by level of competence
          (decreasing classification)
  Classification_vp(In: List_vp_actor_prod; Out: List_vp_class);
  For i = 1 to size(List_vp_class) Do
    // 4th step: Restitution of information batches and the expertise level for
            each viewpoint using the focus activity and level competency.
    Restitution_list_connexion_level(In: List_vp_class;
                                     Out: List_connexion_level_vp);
    // 5th step: Update of the list of information batches and its level of use
    If (i>1) Then
      Optimize_list_connexion_level(In: List_connexion_level_vp;
                                    Out: List_lev_vp_prebious);
    Else
      List_con_lev_vp_previous = List_connexion_level_vp
    End_If
  End_For
  List_connexion_batch_info = List_connexion_level_vp;
End
```

**Table 1: Algorithm of information by actor's viewpoints.**

## 4- Example

To illustrate the proposed approach, we present an example about a closed pack cyclone vessel. In the next sections, we present the three models, product structure (or decomposition), process development, and the development of the supply-chain organization. Figure 9 shows examples of two viewpoints of the same actor of the information restitution.

### 4.1 – Product structure example

Figure 6 shows a model representing the decomposition of a closed pack cyclone vessel into 18 components. The product architecture and the interactions between the components are documented and detailed in this model (not presented in this paper). 38 interactions were identified according to a classification related to space, energy, material, and other information.

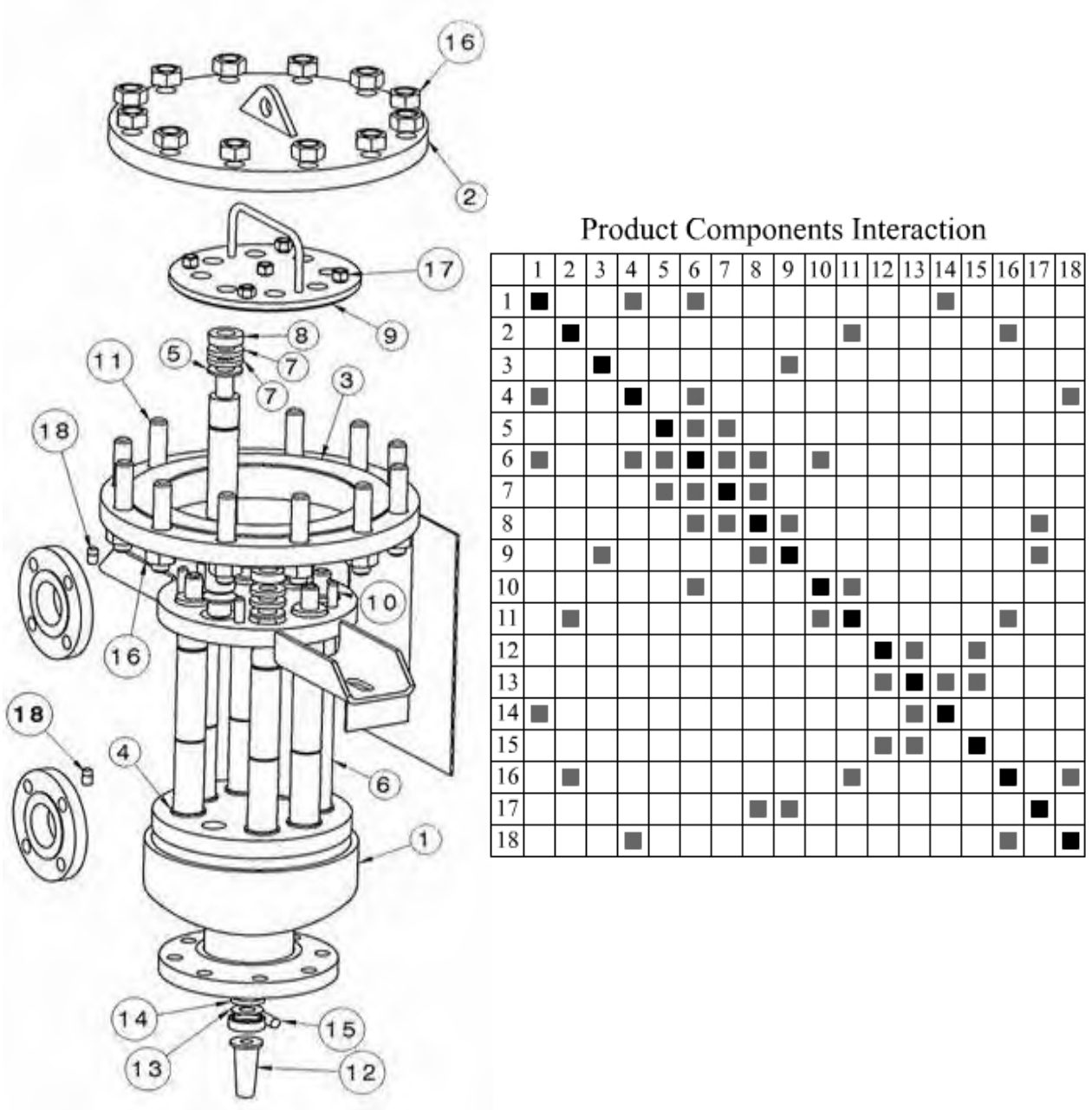

**Figure 6: Cyclone vessel structure.**

### 4.2 – Development process example

Figure 7 shows a tree illustrating the procedure followed by a vessel manufacturer to determine the feasible layout of the customer requirements of the closed pack cyclone vessel. This is based on a digital model using CAD solid models. Interactions in this type of model represent flows of information and data between the tasks of the activities (in this example, we consider that every process has only one activity).

### 4.3 – Supply-chain organisation example

Figure 8 shows the decomposition of the supply-chain organization used to develop a new cyclone vessel project. The organization involves 3 teams; each team has a responsibility for a major component. The given matrix depicts the interactions across the 3 teams in terms of frequency of their required collaboration.

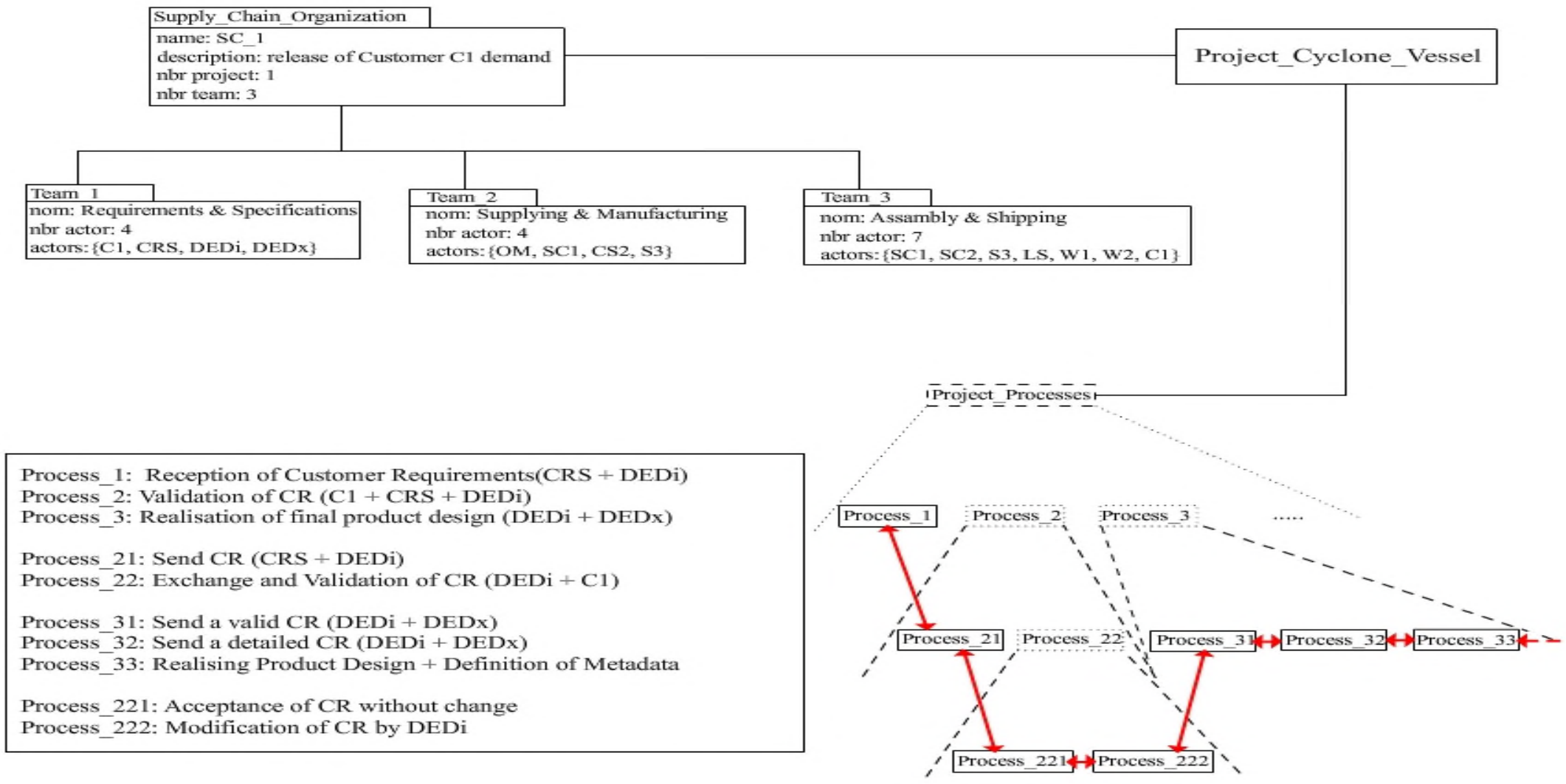

**Figure 7: Part of the process development.**

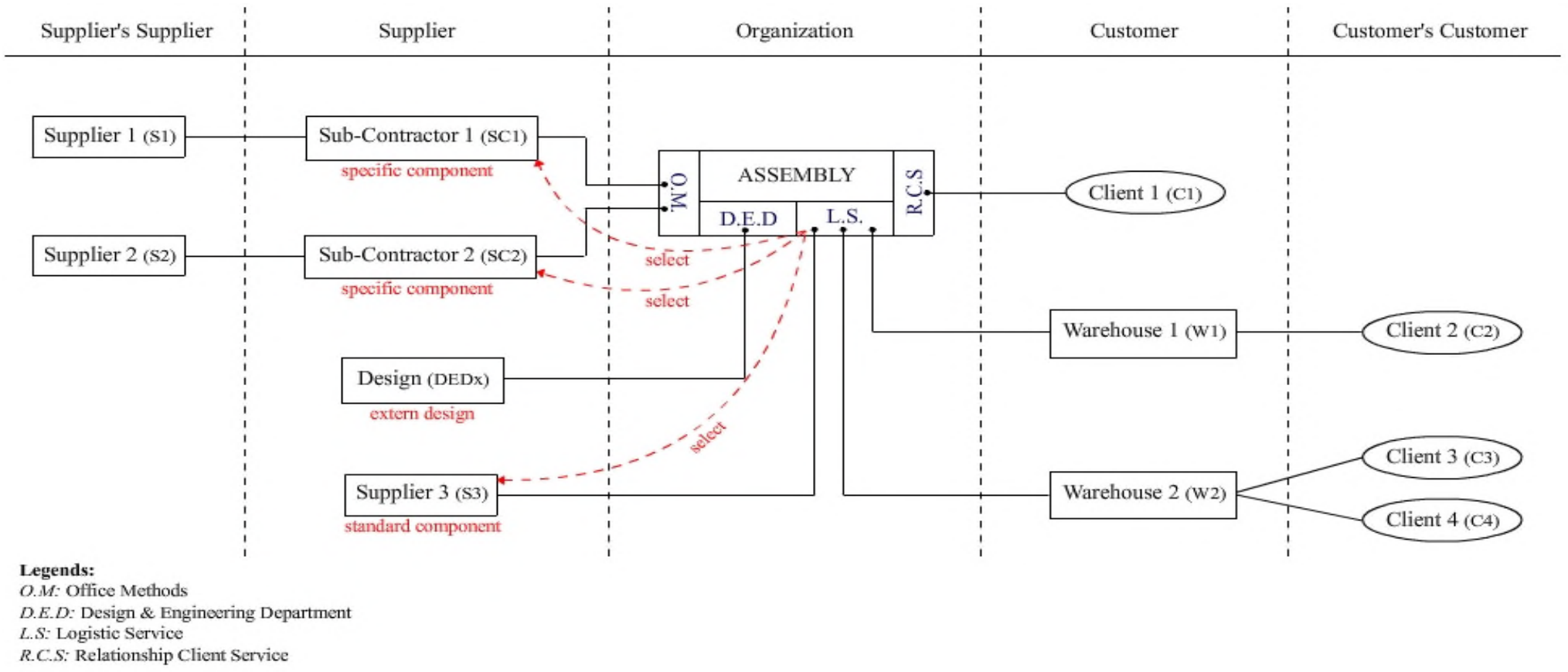

**Figure 8: Development of the supply-chain organization.**

### 4.4 – Information restitution by viewpoint approach

Let's take the example of an actor "ActorX" integrating the supply chain organization into the team 1 as *an external designer*. He has two focuses on the product "cyclone vessel", the first as **shape global design**, and the second as **mechanical design** (Fig. 9). The actor's objectives are related respectively to the activity on geometry and mechanic tasks. To retrieve the adequate information for the actor "ActorX", we need to filter and classify the information following his interests. After this step, the framework compares the information of each viewpoint (Tab. 2) and gives the sets of adequate information to the actor following his focuses on the product and his activities in the project. Based on the level's batch definition, the system regroups the batches with high-level hierarchy, and retrieves the information which is more adequate to the actor according to his focuses and activities [GB1].

Figure 10 shows part of the generated XML file of the retrieved information, the actor "ActorX" can use/reuse this information.

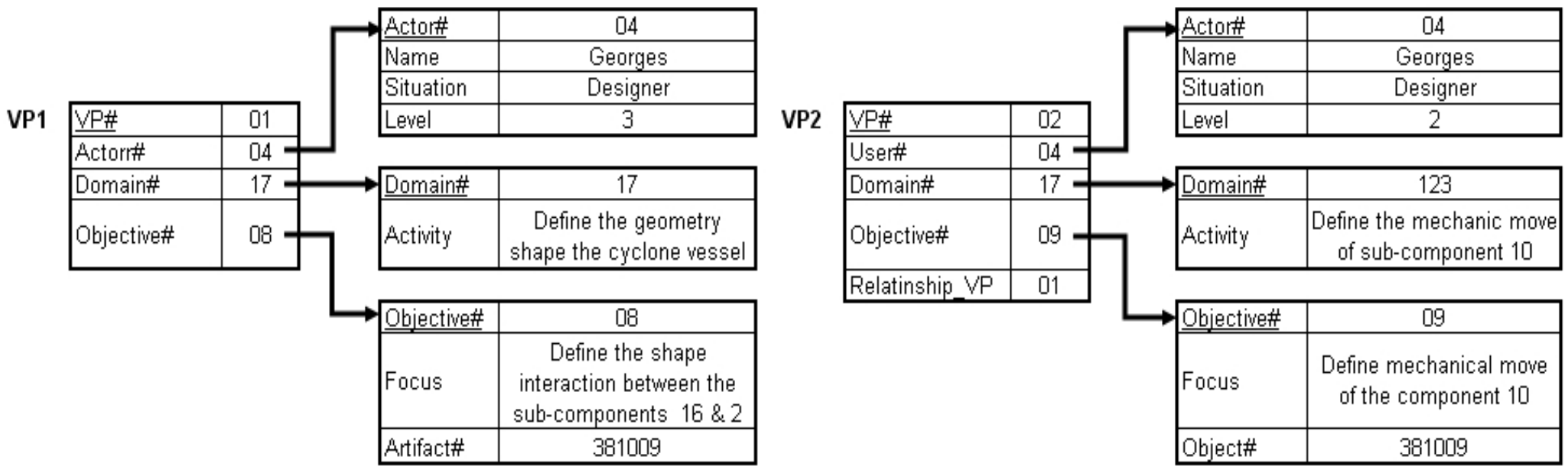

**Figure 9: ActorX viewpoints.**

| VP# = 01 | VP# = 02 |
|---|---|
| Batch (Level):<br><br>Artifact (1) : Standard Information about the product<br>Function (2) : Different function of the final-product and sub-artifact<br>Behavior (2) : Different behaviors of the final-product and sub-artifact in relation with their respective functions<br>Flows (2) : Different flows of the final-product functions<br>Geometry-Form (1) : All information about the detailed-geometry with a CAD model<br>Sub-Artifact (2) : Different information about the second level of direct-component assembly<br>Assembly (2) : The relationship with Sub-Artifacts<br>Constraints (1): All constraints of the product (design, assembly…)<br>Requirements (2) : Requirements about the product and different phases of its lifecycle<br>Group (1) : All information about actors of collaboration-team<br>... | Batch (Level):<br><br>Mechanic (1) : All information about the mechanic application of the product (see activity of VP09 and VP08)<br>Artifact (2) : Standard Information about the product<br>Function (2) : Different function of the final-product and sub-artifact<br>Behavior (2) : Different behaviors of the final-product and sub-artifact in relation with their respective functions<br>Flows (3) : Different flows of the final-product functions<br>Geometry-Form (2) : Different information about the geometry with a CAD model<br>Sub-Artifact (3) : Different information about the first level of direct-component assembly<br>Assembly (3) The relationship with Sub-Artifacts<br>Constraints (1) : All constraints design of the product<br>Requirements (3) : Requirements about the product and different phases of its lifecycle<br>Group (1) : All information about the actors in the group<br>… |

**Table 2: Results after the information filtering by the actor's viewpoints.**

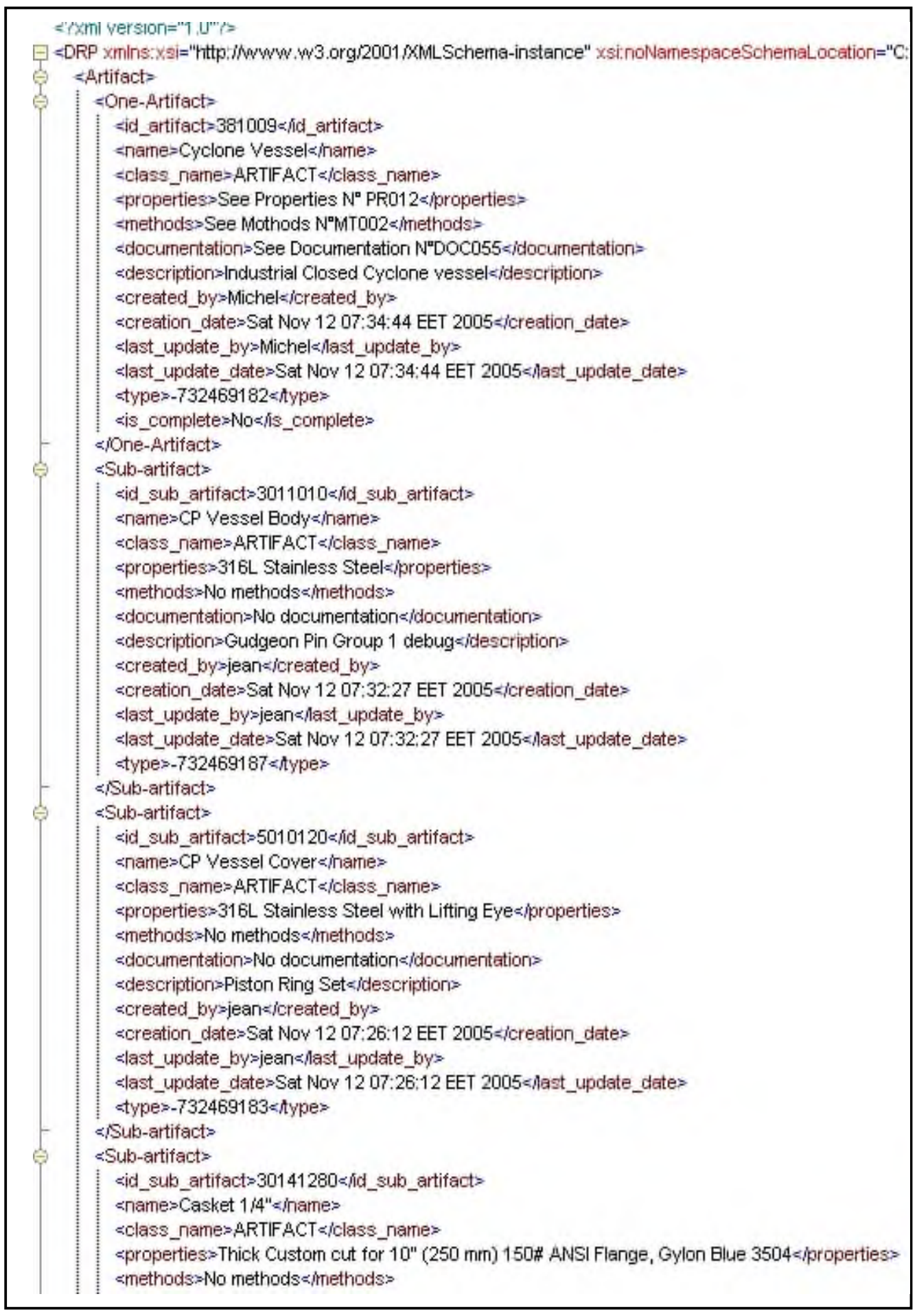

```
<?xml version="1.0"?>
<DRP xmlns:xsi="http://www.w3.org/2001/XMLSchema-instance" xsi:noNamespaceSchemaLocation="C:
  <Artifact>
    <One-Artifact>
      <id_artifact>381009</id_artifact>
      <name>Cyclone Vessel</name>
      <class_name>ARTIFACT</class_name>
      <properties>See Properties N° PR012</properties>
      <methods>See Mothods N°MT002</methods>
      <documentation>See Documentation N°DOC055</documentation>
      <description>Industrial Closed Cyclone vessel</description>
      <created_by>Michel</created_by>
      <creation_date>Sat Nov 12 07:34:44 EET 2005</creation_date>
      <last_update_by>Michel</last_update_by>
      <last_update_date>Sat Nov 12 07:34:44 EET 2005</last_update_date>
      <type>-732469182</type>
      <is_complete>No</is_complete>
    </One-Artifact>
    <Sub-artifact>
      <id_sub_artifact>3011010</id_sub_artifact>
      <name>CP Vessel Body</name>
      <class_name>ARTIFACT</class_name>
      <properties>316L Stainless Steel</properties>
      <methods>No methods</methods>
      <documentation>No documentation</documentation>
      <description>Gudgeon Pin Group 1 debug</description>
      <created_by>jean</created_by>
      <creation_date>Sat Nov 12 07:32:27 EET 2005</creation_date>
      <last_update_by>jean</last_update_by>
      <last_update_date>Sat Nov 12 07:32:27 EET 2005</last_update_date>
      <type>-732469187</type>
    </Sub-artifact>
    <Sub-artifact>
      <id_sub_artifact>5010120</id_sub_artifact>
      <name>CP Vessel Cover</name>
      <class_name>ARTIFACT</class_name>
      <properties>316L Stainless Steel with Lifting Eye</properties>
      <methods>No methods</methods>
      <documentation>No documentation</documentation>
      <description>Piston Ring Set</description>
      <created_by>jean</created_by>
      <creation_date>Sat Nov 12 07:26:12 EET 2005</creation_date>
      <last_update_by>jean</last_update_by>
      <last_update_date>Sat Nov 12 07:26:12 EET 2005</last_update_date>
      <type>-732469183</type>
    </Sub-artifact>
    <Sub-artifact>
      <id_sub_artifact>30141280</id_sub_artifact>
      <name>Casket 1/4"</name>
      <class_name>ARTIFACT</class_name>
      <properties>Thick Custom cut for 10" (250 mm) 150# ANSI Flange, Gylon Blue 3504</properties>
      <methods>No methods</methods>
```

**Figure 10: Part of the generated XML file.**

## 5- Conclusion

Multidisciplinary collaboration is a complex domain, in which all the actors need to exchange and share product and process information. In fact, product information generated by each actor is communicated to all actors in order to integrate them in a shared representation. Knowledge about actors' preferences, methods, etc. and about actors' focuses, constraints, objectives, etc. must be taken into account to manage information extraction.

In Multidisciplinary collaboration, the use of viewpoint in the structured collaborative product development shows how the viewpoint notion can provide real help in the extraction, treatment and consulting of adequate product/process information. The proposed viewpoint description and multi-level management approach aim to structure the actor's focuses in multidisciplinary collaboration thanks to a more accurate characterization of the viewpoints, which allows an intelligent indexation of the product/process information.

In this paper, we have presented our Product, Process and Organisation Models, from the PPCO model, developed in previous studies, the Collaboration Model, is not presented in this paper, but it defines the different strategies/methods of collaboration between the actors in-side the Supply Chain and in the different activities of Project Processes. The originalities of the PPCO model are: i) the integration of the Supply Chain world into the creation of the collaborative product information during the product lifecycle, ii) the opening of the product metadata model, which can integrate any existing product data model, like as the Core Product Model [SF1], and iii) the inter-connection of the model with the viewpoint approach to facilitate the information extraction and exchange between actors.

## 6- References

**[BN1]** Bronsvoort WF., Noort A., Van Den Berg E., Hoek GFM. Product development with multiple-view feature modelling. IFIP, in proceedings of the conference on Feature Modeling and Advanced Design-for-the-Lifecycle Systems, Valenciennes, France, June 2001.

**[BW1]** Bullinger HJ., Warschat J., Fischer D. Rapid product development – an overview. International Journal on Computer-industry, 42 (2): 99-108, 2000.

**[E1]** Easterbrook S. Domain Modelling with Hierarchies of alternative viewpoints. In proceedings of IEEE International Symposium on Requirements Engineering, January 4-6, San Diego, California, 1993.

**[G1]** Garlan D. Views for Tools in Integrated Environments. In proceedings of TOOLS'87, pp. 313-343, 1987.

**[GB1]** Geryville H., Bouras A., Ouzrout Y., Sapidis N. Collaborative product and process model: Multiple

viewpoints approach. The 12th International Conference on Concurrent Enterprising, , pp. 391-398, Milan - Italy, 26-28 June 2006.

**[GR1]** Gzara L., Rieu D., Tollenaere M. Product information systems engineering: an approach for building product models by reuse of patterns. International Journal on Robotics and Computer Integrated Manufacturing, 19(2): 239-261, 2003.

**[GO1]** Geryville H., Ouzrout Y., Bouras A., Sapidis N. A collaborative framework to exchange and sharing product information within a supply chain context. In Proceeding of IEEE, International Conference on Machine Intelligence, pp. 195-202, 4-7 November 2005.

**[HJ1]** Hoffman CM., Joan-Ariyo R. Distributed maintenance of multiple product views. International Journal on Computer-Aided Design, 32(4): 421-431, 2000.

**[I1]** IEEE Standard 1471. Recommended Practice for Architectural Description of Software-Intensive Systems. IEEE, 2000.

**[I2]** ISO/IEC 10303-203. Industrial automation systems and integration – Product data representation and exchange – Part 203: Application protocol: Configuration controlled design. ISO TC184, SC4, 1994.

**[I3]** ISO/IEC. RM-ODP: Reference Model for Open Distributed Processing. International Standard ISO/IEC 10746-1 to 10746-4", ITU-T Recommendations X.901 to X.904. 1997.

**[K1]** Karacapilidis N. Modelling discourse in collaborative work support systems: a knowledge representation and configuration perspective. International Journal on Knowledge-based Systems, 15(7):413-422,2002.

**[K2]** Kruchten P. Architectural blueprints—the "4+1" view model of software architecture. IEEE Software, 12 (6): 42-50, 1995.

**[NR1]** Nowak P., Rose B., Saint-Marc L., Callot M., Eyanard B., Gzara L., Lombard M. Towards a design process model enabling the integration of product, process and organization. The 5th International Conference on Integrated Design and Manufacturing in Mechanical Engineering, Bath, United-Kingdom, April 5-7 2004.

**[R1]** Ribière M. Using viewpoints and CG for the representation and management of a corporate memory in concurrent engineering. Springer-Velag, ICCS'98, pp. 94-108, 1998.

**[SF1]** Sudarsan R., Fnves SJ., Sriram RD., Wang F. A product information modelling framework for product lifecycle management. International Journal of Computer Aided Design, Vol. 37(13), pp. 1399-1411, Nov. 2005.

**[ST1]** Stuurstraat N., Tolman F. A product modelling approach to building knowledge integration. International Journal on Automation in Construction, 8(3): 269-275, 1999.

**[VT1]** Veeramani D., Tserng HP., Russell JS. Computer-integrated collaborative design and operation in the construction industry. International Journal on Automat-Construct, 7(4): 485-492, 1998.

**[Z1]** Zachman J. The Zachman Framework: a primer for enterprise engineering and manufacturing. Zachman International, 1997. http://www.zifa.com.

**[ZL1]** Zhang WJ. Liu SN. Li Q. Data/knowledge representation of modular robot and its working environment. International Journal in Robotics and Computer-Integrated Manufacturing, 16(2-3):143-159, April 2000.